\documentstyle[aps,preprint]{revtex}
\begin{document}
\draft
\tighten
\preprint{\vbox{Submitted to Physical Review D
                \hfill TIT/HEP-407/NP 
 }}
\title{Two-point correlation function with
pion in QCD sum rules }
\author{Hungchong Kim$^1$ \footnote{E-mail : 
hckim@th.phys.titech.ac.jp},
Su Houng Lee$^{2,3}$\footnote{AvH fellow.}  and
Makoto Oka$^1$ }
\address{$^1$ Department of Physics, Tokyo Institute of Technology, Tokyo 
152-8551, 
Japan \\
$^2$ GSI, Planckstr. 1, D-64291 Darmstadt, Germany \\
$^3$ Department of Physics, Yonsei University, Seoul 120-749, Korea }

\maketitle
\begin{abstract}
Within the framework of the conventional QCD sum rules, we study the
pion two-point correlation function, 
$i\int d^4x e^{iq\cdot x} \langle 0| T J_N(x) {\bar J}_N(0)|\pi(p)\rangle$,
beyond the soft-pion limit.  We construct sum rules from 
the three distinct Dirac 
structures,
$i \gamma_5\not\!p$, 
$i \gamma_5$, $\gamma_5 \sigma_{\mu \nu} {q^\mu p^\nu}$
and study the reliability of each sum rule.
The sum rule from the third structure is found to be 
insensitive to the continuum threshold,
$S_\pi$, and contains relatively small contribution from the 
undetermined single pole which we denote as $b$. 
The sum rule from the 
$i \gamma_5$ structure is very different even though
it contains  similar contributions from $S_\pi$ and $b$ as
the ones coming from the $\gamma_5 \sigma_{\mu \nu} {q^\mu p^\nu}$ structure.
On the other hand, the sum rule from the $i \gamma_5$$\not\!p$ structure 
has strong
dependence on both  $S_\pi$ and $b$, which is clearly in constrast with
the sum rule for $\gamma_5 \sigma_{\mu \nu} {q^\mu p^\nu}$.
We identify the source of the sensitivity for each of the sum rules by 
making specific
models for higher resonance contributions and discuss the implication.  

\end{abstract}

\section{INTRODUCTION}
\label{sec:intro}

Since first introduced by Shifman, Vainshtein and Zakharov~\cite{SVZ},
QCD sum rule has been widely used to study the  properties of the 
hadrons~\cite{qsr}. 
QCD sum rule is a framework which connects a physical parameter to the 
parameters of QCD.  In this framework, a correlation function is 
introduced in terms of  interpolating fields, which are 
 constructed from quark and gluon fields.
Then, the correlation function, 
on the one hand,  is calculated by Wilson's
operator product expansion (OPE) and, on the other hand, its  phenomenological
``ansatz'' is constructed.  A physical quantity of
interest is extracted by matching the two descriptions in the
deep Euclidean region ($q^2 = - \infty$) via  the dispersion relation.
The extracted value therefore should be independent of the possible ansatz in
order to be phyically meaningful.

The two-point correlation function with pion,
\begin{eqnarray}
\Pi (q, p) = i \int d^4 x e^{i q \cdot x} \langle 0 | T[J_N (x) 
{\bar J}_N (0)]| \pi (p) \rangle \ ,
\label{two}
\end{eqnarray}
is often used to calculate the pion-nucleon coupling, $g_{\pi N}$, in QCD sum 
rules~\cite{qsr,hat,krippa}.
Reinders, Rubinstein and Yazaki~\cite{qsr} calculated
$g_{\pi N}$ by retaining only the first nonperturbative term in 
the OPE.  Later Shiomi and Hatsuda (SH)~\cite{hat} improved
the calculation by including higher order terms in the OPE. SH considered
Eq.~(\ref{two})
and evaluated the OPE in the soft-pion limit ($p_\mu \rightarrow 0$).

More recently, Birse and Krippa (BK)~\cite{krippa}  pointed out that
the use of the soft-pion limit does not constitute an independent sum
rule from the nucleon sum rule because in the limit the correlation function
is just a chiral rotation of the nucleon correlation function,
\begin{eqnarray}
\Pi (q) = i \int d^4 x e^{i q \cdot x} \langle 0 | T[J_N (x) 
{\bar J}_N (0)] | 0 \rangle \ .
\label{ntwo}
\end{eqnarray}
Therefore, BK considered the sum rule beyond the soft-pion limit.
However, as we will discuss below,
there seems to be mistakes in their calculation
which can invalidate their conclusions.   
Thus, it is important to re-do their calculation.


In a recent letter~\cite{hung}, we have pointed out that
the previous calculations of the pion-nucleon coupling using
Eq.~(\ref{two}) have dependence on how 
one models the phenomenological side; either using the 
pseudoscalar (PS) or the 
pseudovector (PV) coupling scheme.  The two coupling schemes
are equivalent when the participating nucleons are on-shell but
they are 
not usually when the nucleons are off-shell.
Since, in QCD sum rules, on-shell properties of a particle are 
extracted from the far off-shell point, the extracted $g_{\pi N}$ therefore
could be coupling-scheme dependent. 
Going beyond the
soft-pion limit is found to be also natural in obtaining
$g_{\pi N}$ independent of the PS and PV coupling schemes.
In fact, we have proposed that, beyond the
soft-pion limit, there are three 
distinct Dirac structures, 
(1) $i \gamma_5 \not\!p$,
(2) $i \gamma_5$, (3) $\gamma_5 \sigma_{\mu \nu} {q^\mu p^\nu}$, 
each of which can in principle be used to calculate
$g_{\pi N}$.  The third structure was found to have the common
double pole structure in the phenomenological
side, independent of the PS and PV
coupling schemes. By studying
this structure, we obtained the coupling close to its empirical
value and relatively stable against the uncertainties from QCD parameters.  
Then we ask, can we get similar stable results
from the sum rules constructed from the other Dirac structures ?  
If not, what are the
reasons for the differences ?
In this work, we will try to answer these questions
by studying these three sum rules  and
investigating the reliability of each sum rule.

QCD sum rules could depend on
a specific Dirac structure considered.
This aspect was suggested by  Jin and Tang~\cite{jin2}
in their study of baryon sum rules.
They found that the chiral odd sum rule is more reliable due
to the partial cancellation of the positive and negative-parity
excited baryons in the continuum.  Similarly here
we note that the structure (1) has different chirality from the other
two.  Therefore it will be interesting to
look into these sum rules more closely and see if
similar cancellation occurs for certain sum rules.

The paper is organized as follows. In Section~\ref{sec:qcdsr},
we construct three sum rules from the three different Dirac structures.
The spectral density for the phenomenological side is constructed
from the double pole, the unknown single pole and  the
continuum modeled by a step function. 
We motivate this phenomenological spectral density
in Section~\ref{sec:effective}
by using
some effective Lagrangians for the
transitions, $N \rightarrow N^*$ and $N^* \rightarrow N^*$. 
In Section~\ref{sec:analysis}, we analyze each sum rule and
try to understand the differences from the formalism constructed
in Section~\ref{sec:effective}.  A summary is given in 
Section~\ref{sec:summary}.

\section{QCD sum rules for the two-point correlation function}
\label{sec:qcdsr}

In this section, we formulate three different sum rules for the two-point
correlation function with pion beyond the soft-pion limit.  
For technical simplicity, we consider the correlation function with
charged pion,
\begin{eqnarray}
\Pi (q,p) = i \int d^4 x e^{i q \cdot x} \langle 0 | T[J_p (x) 
{\bar J}_n (0)]| \pi^+ (p) \rangle\ .
\label{two2}
\end{eqnarray}
Here $J_p$ is the proton interpolating field suggested by Ioffe~\cite{ioffe1},
\begin{eqnarray}
J_p = \epsilon_{abc} [ u_a^T C \gamma_\mu u_b ] \gamma_5 \gamma^\mu d_c
\end{eqnarray}
and the neutron interpolating field $J_n$ is obtained by replacing
$(u,d) \rightarrow (d,u)$.  
In the OPE,  we only keep the diquark component of the 
pion wave function and use the vacuum saturation hypothesis
to factor out 
higher dimensional operators in terms of the pion wave function and the 
vacuum expectation value.

The calculation of the correlator, Eq.~(\ref{two2}), in the coordinate
space contains the following diquark component of the pion wave function, 
\begin{eqnarray}
D^{\alpha\beta}_{a a'} \equiv 
\langle 0 | u^\alpha_a (x) {\bar d}^\beta_{a'} (0) | \pi^+ (p) \rangle\ .
\end{eqnarray}
Here, $\alpha$ and $\beta$ are Dirac indices, $a$ and $a'$ are
color indices. 
The other quarks are contracted to form quark propagators.
This diquark component can be written in terms of three
Dirac structures,
\begin{eqnarray}
D^{\alpha\beta}_{a a'} = && {\delta_{a a'} \over 12} 
(\gamma^\mu \gamma_5)^{\alpha \beta}
\langle 0 |  
{\bar d} (0) \gamma_\mu \gamma_5  u (x) | \pi^+ (p) \rangle\ 
+ {\delta_{a a'} \over 12 } (i \gamma_5)^{\alpha \beta} 
\langle 0 |  
{\bar d}(0) i \gamma_5  u (x) | \pi^+ (p) \rangle\nonumber \\ 
&& - {\delta_{a a'} \over 24} (\gamma_5 \sigma^{\mu\nu})^{\alpha\beta}
\langle 0 |  
{\bar d}(0) \gamma_5 \sigma_{\mu\nu}  u (x) | \pi^+ (p) \rangle\ .
\label{dd}
\end{eqnarray}
Each matrix element associated with each Dirac 
structure can be written in terms of pion wave 
function whose first few moments are relatively well known~\cite{bely}.
We will come back to the second matrix element later. For the other two
elements, we need only the normalization of the pion wave functions since
we are doing the calculation up to the first order in $p_\mu$. 
In fact, to leading order in the pion momentum, 
the first and third matrix elements are given as~\cite{bely},
\begin{eqnarray}
\langle 0 | {\bar d} (0) \gamma_\mu \gamma_5  u (x) | \pi^+ (p) \rangle\ 
&=& i \sqrt{2} f_\pi p_\mu + {\rm twist~4~term}\ , \label{d1} \\
\langle 0 |{\bar d}(0) \gamma_5 \sigma_{\mu\nu}  u (x) | \pi^+ (p) \rangle\ 
&=&i \sqrt{2} (p_\mu x_\nu - p_\nu x_\mu) 
{f_\pi m_\pi^2 \over 6 (m_u + m_d)}\ .
\label{d3}
\end{eqnarray}
Here we have suppressed terms higher order in pion momentum.
The factor $\sqrt{2}$ is just an isospin factor. The
twist 4 term in Eq.~(\ref{d1}) comes from the second derivative
term in the short distant expansion of the LHS.   
Note that in Eq.~(\ref{d3}) 
the factor $f_\pi m_\pi^2 / (m_u + m_d)$ can
be written as $-\langle {\bar q} q \rangle / f_\pi$ by making use of
Gell-Mann$-$Oakes$-$Renner relation.  
Although the operator looks gauge 
dependent, it is understood that the fixed point gauge is used throughout 
and the final result is gauge independent.  
It is then interesting to note that 
the LHS of  Eq.~(\ref{d3}) can also be expanded in $x$ such 
that the matrix element that contributes is effectively one with higher
dimension,
\begin{eqnarray}
\langle 0 |{\bar d}(0) \gamma_5 \sigma_{\mu\nu}  D_\alpha  
u (0) | \pi^+ (p) \rangle
 =i \sqrt{2} (p_\mu g_{\alpha \nu} - p_\nu g_{\alpha \mu}) 
{f_\pi m_\pi^2 \over 6 (m_u + m_d)}\ .
\label{su}
\end{eqnarray}

It is now straightforward to calculate the OPE.  For  the 
$i\gamma_5\not\!p$ structure, we obtain
\begin{eqnarray}
\sqrt{2} f_\pi \left [ {q^2 {\rm ln} (-q^2) \over 2 \pi^2 }+
{\delta^2 {\rm ln} (-q^2) \over 2 \pi^2} + 
{\left \langle {\alpha_s \over \pi} {\cal G}^2 
\right \rangle \over 12 q^2} +
{2 \langle {\bar q} q \rangle^2 \over 9 f_\pi^2 q^2} \right ]\ .
\label{bkope}
\end{eqnarray}
The first three terms are obtained by taking the second term in
Eq.~(\ref{dd})\footnote{Note that the second term in Eq.~(\ref{bkope}) has
slightly different coefficient from BK~\cite{krippa}. 
Ref.~\cite{krippa} has the factor 5/9 instead of our factor 1/2.  The 
difference however is small.}, while
the fourth term is obtained by taking the third term in Eq.(\ref{dd})
and  replacing one quark propagator with 
the quark condensate. The fourth term was not taken into account
in the sum rule studied by BK~\cite{krippa} but
its magnitude is about 4 times larger than the third term. So there is
no reason to neglect the fourth term while keeping the third term. 
The second term comes from  the twist-4 element of pion wave function.
According to Novikov {\it et. al}~\cite{nov}, $\delta^2 \sim 0.2$ GeV$^2$.

The phenomenological side for the $i\gamma_5\not\!p$ structure 
obtained by using the pseudoscalar Lagrangian takes the form,
\begin{eqnarray}
-{\sqrt{2} g_{\pi N} \lambda_N^2  m  \over 
(q^2 - m^2 +i \epsilon)[(q-p)^2 - m^2 + i \epsilon]} + \cdot \cdot \cdot\ . 
\label{bkcor}
\end{eqnarray}
The dots include contributions from the continuum as well as from
the unknown
single pole terms .  The latter consists of the
single pole coming from $N \rightarrow N^*$ transition~\cite{ioffe2}.
When the pseudovector Lagrangian is used, there is an additional
single pole coming from $N\rightarrow N$ transition~\cite{hung}. 
These single poles  are not suppressed by the Borel transformation. 
Therefore, interpretation of 
the unknown single pole and possibly the continuum
contain some ambiguity due to the coupling scheme adopted.

In principle, $g_{\pi N}$ has $p^2$ dependence as it 
contains the pion form factor.  As one pion momentum is taken out
by the Dirac structure, $i\gamma_5\not\!p$, we take $p_\mu =0$ 
in the rest of the correlator as we did in the OPE side. 
Then the $p^2$ dependence of
$g_{\pi N}$ can be neglected.  Furthermore, 
after taking out the factor, $i\gamma_5\not\!p$, the rest of the 
correlator is a function of one variable,
$q^2$, and therefore 
the single dispersion relation  in $q^2$ can be invoked in constructing
the sum rule. 
Anyhow, the spectral density can be
written as
\begin{eqnarray}
\rho^{phen} (s) = -\sqrt{2} g_{\pi N} \lambda_N^2  m {d \over ds} 
\delta(s-m^2) + A'~ \delta(s-m^2) + \rho^{ope}(s)~ \theta (s -S_\pi)\ .
\label{bkphen}
\end{eqnarray}
Here the second term comes from the single pole terms whose coefficients
are not known.  
The continuum contribution is parameterized by a step function
which starts from the threshold, $S_\pi$.  
The coefficient of the step function, $\rho^{ope}(s)$,  is 
determined by the duality of QCD. This is basically the imaginary part
of Eq.~(\ref{bkope}) but, because of the continuum threshold, only the first 
two terms in Eq.~(\ref{bkope}) contribute to the coefficient.

The parameterization of the continuum with a step function
is usually adopted in the baryon mass sum rules. This is because
each higher resonance has a single pole structure with a finite width. 
Spectral density obtained by adding up all those single poles
can be effectively represented by a step
function starting from a threshold.  
But in  our case of the correlation function with pion, 
this parameterization for the continuum
could be questionable.  Therefore, it will be useful to construct
the spectral density explicitly for higher resonances  by employing some
effective models for $N^*$ and see if the
parameterization does make sense.
This will be done in the next section.  This will 
eventually help us to understand how each sum rule
based on a different Dirac structure leads to different results.

To construct QCD sum rule for the $i \gamma_5 \not\!p$ structure,
we integrate $\rho^{ope} (s)$ and  $\rho^{phen} (s)$ 
with the Borel weighting factor $e^{-s/M^2}$ and match both sides. More 
specifically, the sum rule equation after the Borel transformation
is given by 
\begin{eqnarray}
\int^\infty_0 ds e^{-s/M^2} [\rho^{ope}(s) - \rho^{phen} (s)] =0\ .
\end{eqnarray}
Using $\rho^{ope}(s)$ obtained from
Eq.~(\ref{bkope}) and $\rho^{phen} (s)$ in 
Eq.~ (\ref{bkphen}), we obtain
\begin{eqnarray}
&&g_{\pi N} \lambda^2_N (1 + A M^2) \nonumber \\
&&= {f_\pi \over m} e^{m^2/M^2} \left [
{E_1 (x_\pi) \over 2 \pi^2} M^6 + {E_0 (x_\pi) \over 2 \pi^2} M^4 \delta^2
+ M^2 \left ( {1 \over 12} 
\left \langle {\alpha_s \over \pi} {\cal G}^2 
\right \rangle  +
{2 \langle {\bar q} q \rangle^2 \over 9 f_\pi^2 }\right ) \right ]\ .
\label{bksum}
\end{eqnarray}
Here $A$ denotes the unknown single pole contribution, which should be
determined by the best fitting method.  
Also $x_\pi = S_\pi/M^2$  and 
$E_n (x) = 1 -(1+x+ \cdot \cdot \cdot + x^n/n!)~ e^{-x}$ .
This expression is crucially different from the corresponding expression
in Ref.~\cite{krippa} where the first, second  and third terms contain 
the factors, $E_2(x_\pi)$,  $E_1(x_\pi)$  and 
$E_0 (x_\pi)$ respectively. Even though we do not understand how such factors 
can be obtained,  
we nevertheless reproduce their figure by using their formula 
in Ref.~\cite{krippa}
and it is shown in 
Fig.~\ref{fig1}~(a)\footnote{In plotting Figs.~\ref{fig1}, 
we did not include the last term involving $\langle {\bar q} q \rangle^2$
in Eq.~(\ref{bksum}) as this term is new in our calculation.}. 
But if Eq.~(\ref{bksum})
is used instead, we get Fig.~\ref{fig1}~(b) using the same parameter set
used in Ref.~\cite{krippa}.  The variation scale
of $g_{\pi N}$ in this figure is clearly different from the one 
in Fig.~\ref{fig1}~(a). Note that some of their parameters are quite
different from ours used in our analysis later part of this work.  
For example, $\delta^2=0.35 $ GeV$^2$ is used
in Ref.~\cite{krippa}, which is quite larger than our value of 0.2 GeV$^2$.

QCD sum rule for the  $\gamma_5 \sigma_{\mu\nu} q^\mu p^\nu$ structure can be 
constructed similarly.
We have constructed the sum rule for this structure in Ref.~\cite{hung}
so here we simply write down the resulting expression, 
\begin{eqnarray}
g_{\pi N} \lambda_N^2  ( 1+ B M^2  )= - {\langle {\bar q} q \rangle \over 
f_\pi} e^{m^2/M^2} \left [ {M^4 E_0 (x_\pi) \over 12 \pi^2 }
+{4 \over 3 } f^2_\pi M^2  + 
\left \langle {\alpha_s \over \pi} {\cal G}^2
\right \rangle 
{1 \over 216 } 
-{m_0^2 f^2_\pi \over 6 }
\right ]\ .
\label{hungsum}
\end{eqnarray}
Here $B$ denotes the contribution from the unknown single pole term.
Note that, since one power of the pion momentum is taken out 
by the factor, $\gamma_5 \sigma_{\mu\nu} q^\mu p^\nu$,
we take the limit $p_\mu = 0$ in the rest of the correlator
as we did in the $i\gamma_5 \not\!p$ case.
In obtaining  the first and third terms in RHS, we have 
used Eq.~(\ref{d3}) while 
the second is  obtained by taking the first 
term in Eq.~(\ref{dd})
for the matrix element $D^{\alpha\beta}_{aa'}$  and replacing one
propagator with the quark condensate. 
The fourth term is also obtained by taking the first term in
Eq.~(\ref{dd}) but in this case other quarks are used to form
the dimension five mixed condensate, 
$\langle {\bar q} g_s \sigma \cdot {\cal G} q \rangle $,
which is usually parameterized in terms of the quark condensate,
$m_0^2 \langle {\bar q} q \rangle $.
We take $m_0^2 \sim 0.8$ GeV$^2$  as obtained from QCD sum rule
calculation~\cite{Ovc}.

Now we construct QCD sum rule for the $i \gamma_5$ structure.
Constructing it beyond the soft-pion limit is more complicated 
as the correlator in phenomenological side has definite dependence
on the coupling schemes. To see this, we expand the correlator
for this structure in $p_\mu$ and write
\begin{equation}
\Pi_0 (q^2) + p\cdot q \Pi_1 (q^2)+p^2 \Pi_2 (q^2) + \cdot \cdot \cdot\ .
\label{expansion}
\end{equation}
Since $p_\mu$ is an external momentum, the correlation function at
each order of $p_\mu$ can be used to construct an independent sum rule.
Within the PS coupling scheme, the phenomenological correlator
up to $p^2$ order is
\begin{eqnarray}
\sqrt{2} g_{\pi N} \lambda_N^2 \left [ 
-{1 \over q^2 - m^2 }
-{ p \cdot q \over
(q^2 - m^2 )^2 } 
+{ p^2  \over
(q^2 - m^2 )^2 } \right ] 
-{\sqrt{2} \lambda_N^2 p^2 \over q^2 - m^2 } {d g_{\pi N} \over d p^2}(p^2 =0)
 \cdot \cdot \cdot\ .
\label{shphen1}
\end{eqnarray}
The dots here represent not only the contribution from higher resonances 
but also terms higher than $p^2$.
The last term is related to the slope of the pion form factor at $p^2 =0$.
Even though this can be absorbed into the unknown single pole term such 
as $A$ or $B$ above, we specify it here since this 
possibility is new.  
This correlator can be compared with the corresponding expression 
in the PV coupling scheme,
\begin{eqnarray}
\sqrt{2} g_{\pi N} \lambda_N^2  {p^2/2 \over
(q^2 - m^2 )^2} + \cdot \cdot \cdot\  .
\label{shphen2}
\end{eqnarray}
Note here that  there are no terms corresponding to $\Pi_0$ and $\Pi_1$.  
No such terms  can be constructed from $N\rightarrow N^*$ or
$N^* \rightarrow N^*$ transitions within the PV scheme.

The single pole in Eq.(\ref{shphen1}) survives in the soft-pion limit,
which has been used by SH~\cite{hat} for their sum rule
calculation of $g_{\pi N}$.  However, if the
phenomenological correlator in PV scheme is used,  
such sum rule cannot be constructed.
Thus, going beyond
the soft-pion limit seems to be natural for the independent determination of
the coupling.
However, similarly for $\Pi_0$ case, a sum rule  can not be
constructed for $\Pi_1$.  For $\Pi_2$, a sum rule can be constructed
either in the PS or PV coupling scheme, but the residue of the
double pole in Eq.~(\ref{shphen2}) is a factor of two smaller
than the corresponding term in Eq.~(\ref{shphen1}).
So the coupling-scheme independence can not be achieved in any of
these sum rules. 
This is true for even higher orders of $p_\mu$.

A sum rule, independent of the coupling schemes, can be constructed
by imposing the kinematical condition,
\begin{equation}
p^2 = 2 p \cdot q\ .
\label{cond}
\end{equation}
With this condition,
the two double pole terms in Eq.~(\ref{shphen1}) can be combined
to yield the same expression as in Eq.~(\ref{shphen2}),
thus providing a sum rule independent of the
coupling schemes.  
This condition comes from the on-shell conditions
for the participating nucleons,
$q^2 = m^2 $ and $(q-p)^2 = m^2$, at which the physical $\pi NN$
coupling should be defined.

The sum rule constructed with the kinematical condition, Eq.~(\ref{cond}),
is equivalent to consider $\Pi_1(q^2)/2+\Pi_2 (q^2)$.
This sum rule seems fine in the PS coupling scheme as
there are nonzero terms corresponding to $\Pi_1$ and $\Pi_2$.
In the OPE, the diquark component contributing to $i \gamma_5$ structure is 
the second element of Eq.~(\ref{dd}) which can be written in terms of twist-3
pion wave function as~\cite{bely}
\begin{eqnarray}
\langle 0 |  
{\bar d}(0) i \gamma_5  u (x) | \pi^+ (p) \rangle =
{\sqrt{2} f_\pi m_\pi^2 \over m_u + m_d} 
\int^1_0 du e^{-i u p\cdot x} \varphi_p (u)\ .
\label{psope}
\end{eqnarray}
The terms linear and quadratic in $p_\mu$ in the RHS 
constitute the OPE correlator for
$\Pi_1$ and $\Pi_2$. 
Therefore, within the PS scheme, $\Pi_1$ and $\Pi_2$ are well 
defined in both sides.

Situation becomes subtle when the PV coupling scheme is employed.
Before the condition of Eq.~(\ref{cond})
is imposed, a sum rule can be
constructed only for $\Pi_2$ as there is no $\Pi_1$ part in the
phenomenological part.
But after the condition, the phenomenological side has only
$\Pi_2^{phen}$ which should be matched with
$\Pi_1^{ope}/2 + \Pi_2^{ope}$.   This seems a little awkward.
Nevertheless, to achieve the independence of the coupling
schemes,  we construct a QCD sum rule for $i\gamma_5$ within
the kinematical condition, Eq.~(\ref{cond}). 
To be consistent with the expansion in the phenomenological side,
we take the terms up to the order $p^2$ in the expansion of Eq.~(\ref{psope}).
Using the parameterization for $\varphi_p (u)$
given in Ref.~\cite{bely},
we obtained up to $p^2$,
\begin{eqnarray}
\langle 0 | 
{\bar d}(0) i \gamma_5  u (x) | \pi^+ (p) \rangle =
{\sqrt{2} f_\pi m_\pi^2 \over m_u + m_d} 
\left ( 1 -
i {1 \over 2} p \cdot x -{0.343 \over 2} (p\cdot x)^2 \right )\ .
\label{psope1}
\end{eqnarray}
A different parameterization given in Ref.~\cite{bely}
changes the numerical factors very slightly .

Using the diquark component of Eq.~(\ref{psope1}), the OPE side
for $\Pi_1/2 + \Pi_2$ is 
calculated straightforwardly. By matching with its phenomenological
counterpart and taking the Borel transformation, we get
\begin{eqnarray}
g_{\pi N} \lambda^2_N (1 + C M^2) = {\langle {\bar q} q \rangle 
\over f_\pi} e^{m^2/M^2} \left [
{0.0785 E_0 (x_\pi) \over \pi^2} M^4
- 0.314\times {1 \over 24} 
\left \langle {\alpha_s \over \pi} {\cal G}^2 
\right \rangle  \right ]\ .
\label{shsum}
\end{eqnarray}
Here $C$ again denotes the unknown single pole term which is not
suppressed by the Borel transformation.
This sum rule is different from the other two sum rules  as its first
term in the OPE is
negative.  Each term contains very small
numerical factors due to the cancellation between $\Pi_1^{ope}$
and $\Pi_2^{ope}$.

Up to now, we have presented three different sum rules from Eq.~(\ref{two}).
All these sum rules, in principle, can be used to determine the
pion-nucleon coupling constant, $g_{\pi N}$.  We will discuss the
reliability of each sum rule below.  An alternative  approach is
to consider the nucleon
correlation function in an external axial field as done in Ref~\cite{bely1}.
The nucleon axial charge, $g_A$, calculated in Ref.~\cite{bely1}, agree
well with experiment. Subsequently, by using the Goldberger-Treiman relation,
$g_{\pi N}$ can be also well determined. 
In the approach by Ref.~\cite{bely1}, sum rules for $g_A -1$ is obtained 
by replacing some part of the OPE with the nucleon mass sum rule.
The connection between the sum rules using Eq.~(\ref{two}) and the ones 
in Ref.~\cite{bely1}
is not clear at this moment.  
An important observation made in Ref.~\cite{bely1} is to note that 
some (dominant) terms of
the OPE correspond to a sum rule with $g_A=1$.  This observation allows 
the construction of the sum rules for $g_A -1$.  In our sum rules, this kind 
of observation is not possible. Also the OPE expression from
Eq.~(\ref{two}) is not simply related to the sum rules
with the external field.  Therefore, Eq.~(\ref{two}) seems to
provide independent sum rules from the ones in Ref.~\cite{bely1}.
In future, however, further study is necessary to clarify the connection 
between these two sets 
of independent sum rules as it might provide important aspects for the
nonperturbative nature of hadrons.

\section{Construction of the unknown single pole and the continuum}
\label{sec:effective}

In this section, we construct the unknown single pole term and  the continuum 
by using effective models for the
higher resonances.  This will provide a better understanding of the
parameterization for the continuum in Eq.~(\ref{bkphen})
and give further insights for the unknown single pole term.
Later, this construction will help us to understand the 
differences between each sum rule based on a different Dirac structure.

There are two possible sources for the unknown single pole term and
the continuum. One is from the transition,
$N\rightarrow N^*$ and the other is from the transition, $N^* \rightarrow N^*$.
Of course, as we pointed out in Ref.~\cite{hung}, there could
be additional single pole of nucleon coming from $N \rightarrow N$, 
which however, in the first order of the pion momentum,
appears only in the sum rule  for the $i\gamma_5 \not\!p$
structure within  the PV coupling
scheme.  First we avoid such possibility by constructing
effective models within the PS coupling scheme.  
Later we will discuss
the case with the PV coupling scheme. 
Moreover, we will discuss  only the two Dirac structures,
$i\gamma_5 \not\!p$ and $\gamma_5 \sigma_{\mu\nu} q^\mu p^\nu$. 
The correlator for the $i\gamma_5$ structure with the kinematical
condition of Eq.~(\ref{cond}) takes almost the same form as
the one for the $\gamma_5 \sigma_{\mu\nu} q^\mu p^\nu$ structure.
A slight difference is the appearance of terms containing the 
derivative of pion form factor as indicated in Eq.~(\ref{shphen1}).
Note that this difference is only specific to the PS coupling scheme.
As the form factor is a smooth function of $p^2$ around $p^2 = 0$, 
this difference is not expected to be crucial.

Within the PS coupling scheme, $N \rightarrow N^*$ contributions to
the correlator, Eq.~(\ref{two2}), can be constructed by
using the effective Lagrangians for the positive ($\psi_+$)
and negative ($\psi_-$) parity resonances,
\begin{eqnarray}
g_{\pi NN_+} {\bar \psi} i \gamma_5 {\mbox{\boldmath $\tau$}} \cdot
{\mbox {\boldmath $\pi$}} \psi_+
&+&
g_{\pi NN_+} {\bar \psi}_+ i \gamma_5 {\mbox{\boldmath $\tau$}} \cdot
{\mbox {\boldmath $\pi$}} \psi \nonumber\ ,\\
g_{\pi NN_-} {\bar \psi} i {\mbox{\boldmath $\tau$}} \cdot
{\mbox {\boldmath $\pi$}} \psi_-
&-&
g_{\pi NN_-} {\bar \psi}_- i {\mbox{\boldmath $\tau$}} \cdot
{\mbox {\boldmath $\pi$}} \psi\ .
\end{eqnarray}
The nucleon field is denoted by $\psi$ here.
These terms contribute to the correlator because
the nucleon interpolating field can couple to the positive and negative
parity resonances via,
\begin{eqnarray}
\langle 0 | J_N | N_+ (k, s) \rangle = \lambda_+ U(k,s)\;; \quad
\langle 0 | J_N | N_- (k, s) \rangle = \lambda_- \gamma_5 U(k,s)\ ,
\end{eqnarray}
where $U(k,s)$ denotes the baryon Dirac spinor and $\lambda_{\pm}$
indicates the coupling strength of the interpolating field to
each resonance with specified parity.

The $\gamma_5 \sigma_{\mu\nu} q^\mu p^\nu$ structure of the correlator
takes the form,
\begin{eqnarray}
{2 \lambda_N \lambda_- g_{\pi NN_-} \over
(q^2 - m^2)(q^2 - m_-^2)} + 
{2 \lambda_N \lambda_+ g_{\pi NN_+} \over
(q^2 - m^2)(q^2 - m_+^2)}\ ,
\end{eqnarray}
which can be compared with the $i\gamma_5 \not\!p$ structure 
\begin{eqnarray}
{2  \lambda_N \lambda_- g_{\pi NN_-} (m_- - m)\over
(q^2 - m^2)(q^2 - m_-^2)} - 
{2  \lambda_N \lambda_+ g_{\pi NN_+} (m_+ + m )\over
(q^2 - m^2)(q^2 - m_+^2)}\ .
\end{eqnarray}
By separating as
\begin{eqnarray}
{1 \over (q^2 - m^2)(q^2 - m_{\pm}^2)} \rightarrow 
-{1 \over m_{\pm}^2 - m^2} \left [ {1 \over q^2 - m^2} - 
{1 \over q^2 - m_{\pm}^2} \right ]
\label{sepa}
\end{eqnarray}
we can see that the transitions, $N\rightarrow N^*$, involve the two single
poles, one with the nucleon pole and the other with the
resonance pole. The former constitutes the unknown single pole as it 
involves the undetermined parameters, $\lambda_{\pm}$ and $g_{\pi NN_{\pm}}$.
In the latter,
the finite width of the resonances can be incorporated 
by replacing $m_{\pm} \rightarrow
m_{\pm} -i \Gamma_{\pm}/2$ in the denominator. Then when it is  
combined with other such single poles from
higher resonances, it produces the spectral density which can be parameterized
by a step function as written in Eq.~(\ref{bkphen}). This also implies 
that the continuum threshold, $S_\pi$, does not need to be different from
the one appearing in the usual nucleon sum rule.

It is now easy to obtain the spectral density for the two Dirac structures
by incorporating the decay width of the resonances.
For the $i \gamma_5 \not\!p$ structure, we have
\begin{eqnarray}
\rho_S (s)&=& 2 \left (-{\lambda_+ g_{\pi NN_+} \over m_+ -m} 
+ {\lambda_- g_{\pi NN_-} \over m_- +m}  \right ) \lambda_N \delta (s-m^2)
\nonumber \\
&+&{2 \lambda_N \lambda_+ g_{\pi NN_+} \over m_+ - m} G(s,m_+)
-{2 \lambda_N \lambda_- g_{\pi NN_-} \over m_- + m} G(s,m_-) 
\label{spec1}
\end{eqnarray}
where
\begin{eqnarray}
G(s,m_{\pm}) = {1 \over \pi} 
{m_{\pm} \Gamma_{\pm} \over (s-m_{\pm}^2)^2 + m_{\pm}^2 \Gamma_{\pm}^2}\ .
\end{eqnarray}
Note that the contribution from the positive-parity resonance is enhanced
by the factor $1/(m_+ - m)$ while the one from the negative-parity 
resonance is
suppressed by the factor $1/(m_- + m)$ .
Similarly for the $\gamma_5 \sigma_{\mu\nu} q^\mu p^\nu$ structure,
we obtain
\begin{eqnarray}
\rho_S (s)&=& 2 \left ({\lambda_+ g_{\pi NN_+} \over m_+^2 -m^2}
+ {\lambda_- g_{\pi NN_-} \over m_-^2 -m^2}  \right )
\lambda_N \delta (s-m^2)
\nonumber \\
&-&{2 \lambda_N \lambda_+ g_{\pi NN_+} \over m_+^2 - m^2} G(s,m_+)
-{2 \lambda_N \lambda_- g_{\pi NN_-} \over m_-^2 - m^2} G(s,m_-)\ .
\label{spec2}
\end{eqnarray}
Note that the superficial relative sign between
the positive- and negative-parity resonances 
are opposite to that in Eq.~(\ref{spec1}).  It means,
depending on the relative sign between $\lambda_+ g_{\pi NN_+}$
and $\lambda_- g_{\pi NN_-}$,  the two contributions add up
in one case or cancel each other in the other case.  
In other words, we can say something about the coefficients 
of $\delta (s - m^2)$ and $G(s,m_{\pm})$, 
by studying the sensitivity of the sum rules
to the continuum or to the single pole.

Additional contribution to the continuum may come from 
$N^*\rightarrow N^*$ transitions. For the off-diagonal transitions 
between two parities,
$N_+ \rightarrow N_- $ and $N_- \rightarrow N_+$,
we use the effective Lagrangians,
\begin{eqnarray}
g_{\pi N_+N_-} {\bar \psi}_+ i {\mbox{\boldmath $\tau$}} \cdot
{\mbox {\boldmath $\pi$}} \psi_-
&-&
g_{\pi N_+N_-} {\bar \psi}_- i {\mbox{\boldmath $\tau$}} \cdot
{\mbox {\boldmath $\pi$}} \psi_+\ ,
\end{eqnarray}
to construct the correlator.
These off-diagonal transitions lead to the spectral density of
\begin{eqnarray}
\rho_{OD} (s) \propto \lambda_- \lambda_+ [ G(s, m_+) - G(s, m_-)]\ ,
\end{eqnarray}
which is therefore suppressed by the cancellation between the two parity 
resonances.

For the diagonal transitions,
$N_+ \rightarrow N_+ $ and $N_- \rightarrow N_-$,
we use
the effective Lagrangians,
\begin{eqnarray}
g_{\pi N_+N_+} {\bar \psi}_+ i \gamma_5 {\mbox{\boldmath $\tau$}} \cdot
{\mbox {\boldmath $\pi$}} \psi_+\;; \quad
g_{\pi N_-N_-} {\bar \psi}_- i \gamma_5 {\mbox{\boldmath $\tau$}} \cdot
{\mbox {\boldmath $\pi$}} \psi_- \ .
\end{eqnarray}
These diagonal transitions produce only the double pole for the correlator,  
$1/(q^2 - m^2_{\pm} + i m_{\pm} \Gamma_{\pm} )^2$, 
which is then translated into the spectral density, 
\begin{eqnarray}
\rho_D (s) \sim \cases { -m_{\pm} g_{\pi N_{\pm} N_{\pm}} 
\lambda_{\pm}^2 {d \over ds} G(s,m_{\pm})~~{\rm for}~~i\gamma_5 \not\!p \cr
\cr
 \pm g_{\pi N_{\pm} N_{\pm}} 
\lambda_{\pm}^2 {d \over ds} G(s,m_{\pm})~~{\rm for}~~\gamma_5 
\sigma_{\mu\nu} q^\mu p^\nu\ .\cr }
\label{specd}
\end{eqnarray}
First note that, because of the derivative, each spectral density 
has a node at $s=m^2_{\pm}$,
positive below the resonance and negative above the
resonance.  
Then under the integration over $s$, 
the spectral
density from the double pole
is partially  canceled, leaving attenuated contribution 
coming from the $s$ dependent Borel weight. 
Indeed, one can numerically check that, for the Roper resonance, 
$\int ds e^{-s/M^2} G(s,m_+)$ is always larger 
than $\int ds e^{-s/M^2} dG(s,m_+)/ds$ for $M^2 \ge 0.7 $ GeV$^2$ and
the cancellation is more effective as $M^2$ increases. 
In general, the continuum contributes more to a sum rule for larger $M^2$. 
Hence the double pole is more suppressed than the single pole
in the region where the continuum is large. 
Further suppression of the double pole continuum can be observed,
for example, by comparing the first equation of Eq.~(\ref{specd}) with
Eq.~(\ref{spec1}).
Even if one assumes 
\footnote{
The nucleon interpolating field, $J_N$, is constructed such that
it couples strongly to the nucleon but weakly to excited states.
Therefore, $\lambda_{+}$ is expected to be smaller than
$\lambda_N$.   
This assumption, therefore, may be regarded as assuming strong  
coupling to the excited baryon.}
that $g_{\pi NN_+} \lambda_N \lambda_+ 
\sim g_{\pi N_+ N_+} \lambda_+^2$
,  then Eq.~(\ref{spec1})
has the enhancing factor of $1/(m_+ - m)$ while the first equation in
Eq.~(\ref{specd})
contains only $m_+$. Thus, the double pole contribution is
much suppressed than the single pole, which can
be checked also from numerical calculations.  The similar suppression 
can be expected for the second equation in Eq.~(\ref{specd}). 
Therefore, we expect that  the continuum mainly comes from the 
single pole of
$1/(q^2 -m^2_{\pm} + i m_{\pm} \Gamma_{\pm})$ which is generated
only from the $N \rightarrow N^*$ transitions.
This will justify the ``step-like'' parameterization of the
continuum as given in Eq.~(\ref{bkphen}).

Now we discuss the case with the PV coupling scheme. 
We use the following Lagrangians
\begin{eqnarray}
&&{g_{\pi N_+N_+}\over 2 m_+}
{\bar \psi}_+ \gamma_5 \gamma_\mu {\mbox{\boldmath $\tau$}} \cdot
\partial^\mu {\mbox {\boldmath $\pi$}} \psi_+
\;; \quad
{g_{\pi N_-N_-}\over 2 m_-}
{\bar \psi}_- \gamma_5 \gamma_\mu {\mbox{\boldmath $\tau$}} \cdot
\partial^\mu {\mbox {\boldmath $\pi$}} \psi_-\ , 
\nonumber \\
&&{g_{\pi NN_+}\over m + m_+}
{\bar \psi}_+ \gamma_5 \gamma_\mu {\mbox{\boldmath $\tau$}} \cdot
\partial^\mu {\mbox {\boldmath $\pi$}} \psi + (H. C.)\ ,
\nonumber \\
&&{g_{\pi NN_-}\over m_- - m}
{\bar \psi}_- \gamma_5 \gamma_\mu {\mbox{\boldmath $\tau$}} \cdot
\partial^\mu {\mbox {\boldmath $\pi$}} \psi + (H. C.)\ ,
\nonumber \\
&&{g_{\pi N_+N_-}\over m_- - m_+}
{\bar \psi}_- \gamma_5 \gamma_\mu {\mbox{\boldmath $\tau$}} \cdot
\partial^\mu {\mbox {\boldmath $\pi$}} \psi_+ + (H. C.)\ .
\end{eqnarray}
These effective Lagrangians in the PV scheme 
are constructed such that the action is the
same as the PS case when the resonances are on-shell .
In this case, complications arise from the possible single pole term 
coming from $N \rightarrow N$~\cite{hung} contribution which was 
absent in the PS scheme. This also means that
there could be additional single poles coming from
$N \rightarrow N^*$ and $N^* \rightarrow N^*$ transitions.  
Note, 
this kind of complication arises only in the $i\gamma_5\not\! p$
case. That is, for the $\gamma_5 \sigma_{\mu\nu} q^\mu p^\nu$
case, we have the same spectral density as given in Eq.~(\ref{spec2}).

As we mentioned above, because the double pole type contribution, 
$1/(q^2 - m^2_{\pm} + i m_{\pm} \Gamma_{\pm} )^2$, 
to the continuum is suppressed,  only single poles are important in 
constructing the spectral density for the unknown single pole and
the ``step-like'' continuum.
To construct the single poles, we consider all possibilities,
$N\rightarrow N$, $N\rightarrow N^*$ and $N^*\rightarrow N^*$.
The coefficient of
$\lambda_N \delta(s -m^2)$, namely the unknown single pole term for
the $i\gamma_5\not\! p$ structure,
can be collected from $N\rightarrow N$ and $N\rightarrow N^*$
transitions, 
\begin{eqnarray}
 {g_{\pi N} \lambda_N \over 2 m} -
2m \left ({\lambda_+ g_{\pi NN_+} \over m_+^2 -m^2}
+ {\lambda_- g_{\pi NN_-} \over m_-^2 -m^2}  \right )\ .
\label{vspec1}
\end{eqnarray}
Compared with the corresponding term in Eq.~(\ref{spec2}), 
this term is differed by 
the first term associated with $N\rightarrow N$. The second
and third terms are the same except for the overall factor,
$-2 m$.

Also the continuum contributions are collected from the terms
containing $1/(q^2 -m_{\pm}^2)$ in the correlator. 
We thus obtain the spectral density for the continuum,
\begin{eqnarray}
&&\left ( g_{\pi N_+ N_+} {\lambda_+^2 \over 2 m_+} + 
g_{\pi N N_+} {2 m_+ \lambda_N \lambda_+ \over m^2_+ - m^2} -
g_{\pi N_+ N_-} {2 m_+ \lambda_+ \lambda_- \over m^2_- - m^2_+}
\right )
G(s, m_+)  \nonumber\ \\
&+&
\left ( g_{\pi N_- N_-} {\lambda_-^2 \over 2 m_-} - 
g_{\pi N N_-} {2 m_- \lambda_N \lambda_- \over m^2_- - m^2} -
g_{\pi N_+ N_-} {2 m_- \lambda_+ \lambda_- \over m^2_- - m^2_+}
\right )
G(s, m_-)\ .
\label{vspec2}
\end{eqnarray}

\section{Reliability of QCD sum rules and possible interpretation}
\label{sec:analysis}

In section~\ref{sec:qcdsr}, we have constructed three sum rules, each for 
the $i\gamma_5\not\!p$, the $i\gamma_5$ and the 
 $\gamma_5 \sigma_{\mu\nu} q^\mu p^\nu$
structures beyond the soft-pion limit. 
Ideally, all
three sum rules should yield 
the same result  for $g_{\pi N}$.
In reality, each sum rule could have uncertainties due to the 
truncation in the
OPE side or large contributions from the continuum. Therefore, depending
on Dirac structures, there could be large or small 
 uncertainties in the determination of the physical parameter.  
This can be checked by looking into the Borel curves
and seeing whether or not they are stable  functions of the Borel mass.  In
the QCD sum rules for baryon masses, the ratio of two different sum rules
is usually taken in extracting a physical mass 
without explicitly checking the stability of each sum rule. 
As pointed out by Jin and Tang~\cite{jin2}, this  could
be dangerous.  In this section, we will demonstrate this issue further by
considering three sum rules provided in section~\ref{sec:qcdsr}.

In Eqs. (\ref{bksum}), (\ref{shsum}) and (\ref{hungsum}), LHS can
be written in the form, $c + bM^2$. The parameter $c$ denotes  the same
quantity, {\it i.e.} $g_{\pi N} \lambda_N^2$,
but $b$ could be  different in
each sum rule.  We can determine $c$ and $b$
by fitting RHS by a straight line within the appropriately chosen
Borel window.  Usually, the maximum Borel mass is determined by restricting 
the continuum contribution to be less than, say, 30 $\sim$ 40 \% of the
first term of the OPE and the minimum Borel mass is chosen by
restricting the highest dimensional term of the OPE to be less than,
say 10 $\sim$ 20 \% of the total OPE.  These criteria lead to the
Borel window centered around the Borel mass $M^2 \sim 1 $ GeV$^2$. 
Further notice that $c$ determined in this way
does not depend on the PS and PV coupling schemes while the interpretation
of $b$ could be scheme-dependent.

In the analysis below, we use the following standard 
values for the QCD parameters,
\begin{eqnarray}
&&\langle {\bar q} q\rangle = -(0.23~{\rm GeV})^3\;; \quad  
\left \langle {\alpha_s \over \pi} {\cal G}^2
\right \rangle = (0.33~{\rm GeV})^4 \nonumber \ ,\\
&&\delta^2 = 0.2~{\rm GeV}^2\;; \quad m_0^2 = 0.8~{\rm GeV}^2\ .
\end{eqnarray}
Uncertainties in these parameters do not significantly
change our discussion below.
For the nucleon mass $m$ and the pion decay constant $f_\pi$, we
use their physical values, $m=0.94$ GeV and $f_\pi = 0.093$ GeV.

In Figure~\ref{fig2}~(a), we plot the Borel curves obtained from
Eqs.(\ref{bksum}), (\ref{shsum}) and (\ref{hungsum}). 
The thick
solid line is from Eq.(\ref{hungsum}),  the thick dot-dashed line from
Eq.(\ref{shsum}) and the thick dashed line from Eq.(\ref{bksum}).
In all three curves, we use $S_\pi=2.07$ GeV$^2$ corresponding to
the mass squared of the Roper resonance.  To check
the sensitivity on $S_\pi$, we have increased the continuum threshold
by 0.5 GeV$^2$  and plotted  in the same figure denoted by respective
thin lines.

In extracting some physical values, one has to fit the curves within
the appropriate Borel window using the function $c + b M^2$.  
The unknown single pole term, $b$, is represented by the slope of
each Borel curve.  The intersection of the best fitting curve with
the vertical axis gives the value of $c$. Figure~\ref{fig2}~(b)
shows the best fitting curves within the Borel window,
$0.8 \le M^2 \le 1.2 $ GeV$^2$.  This window is chosen 
following the criteria mentioned above.  But as the Borel curves
are almost linear around $M^2 \sim 1 $ GeV$^2$, the qualitative aspect
of our results does not change significantly even if we use the slightly
different window.  

The $\gamma_5 \sigma_{\mu\nu} q^\mu p^\nu$ sum rule yields 
$c \sim 0.00308$ GeV$^6$.
To determine $g_{\pi N}$, 
the unknown parameter $\lambda_N$ needs to be eliminated by
combining with the nucleon odd sum rule~\cite{hung}.  
According to the analysis in Ref.~\cite{hung}, 
this sum rule yields $g_{\pi N} \sim 10 $ relatively close to 
its empirical value. 
As can be seen from the thin solid curve which is almost indistinguishable from
the thick solid curve 
in Fig.~\ref{fig2}~(b), this result
is not sensitive to the continuum threshold, $S_\pi$. 
Also note from Table~\ref{tab} that the unknown single pole term
represented by $b$ is relatively small in this sum rule.

The result from the $i\gamma_5$ sum rule is $c \sim -0.0003 $ GeV$^6$,
which is
obtained from linearly fitting the thick dot-dashed curve 
in Fig.~\ref{fig2}~(a).
Even though the thin dot-dashed curve is almost indistinguishable from the
thick dot-dashed curve, the best fitting value for $c$  
with $S_\pi =2.57 $ GeV$^2$ is about 50 \% smaller than the one with
$S_\pi =2.07 $ GeV$^2$.  This is because the total OPE strength of
this sum rule is very small. 
The negative value of $c$ indicates that $g_{\pi N}$ is negative.
Also the magnitude of this  is about a factor of ten smaller than the 
corresponding value from the $\gamma_5 \sigma_{\mu\nu} q^\mu p^\nu$ sum rule.
When this result is combined with the nucleon odd sum rule, then
the extracted $\pi NN$ coupling would be a lot smaller than 
its empirical value and therefore it can not be acceptable as a reasonable
prediction. 
As we discussed in Section~\ref{sec:qcdsr},
the problem might due to the kinematical condition, Eq.~(\ref{cond}).
Though we have introduced this condition in order to achieve the
independence from the coupling scheme employed, this condition
inevitably combines two independent sum rules, $\Pi_1$ and $\Pi_2$
in Eq.~(\ref{expansion}), which reduces the OPE strength.  
This reduction makes the $i\gamma_5$ sum rule less reliable because of the
cancellation of the main terms.
Nevertheless, this study shows that one could get a totally different
result depending on how the sum rule is constructed.

For the $i\gamma_5\not\!p$ sum rule,
the Borel curve around $M^2 \sim 1 $ GeV$^2$ is almost a linear
function of $M^2$.  By linearly fitting the thick dashed curve 
($S_\pi= 2.07$ GeV$^2$),
we get $c\sim -0.00022$ GeV$^6$.  
But with using $S_\pi= 2.57$ GeV$^2$,  we obtain 
$c\sim -0.0023$ GeV$^6$, a factor
of ten larger in magnitude.
Thus, there is a strong sensitivity on $S_\pi$ which changes
the result substantially. Again $c$  is
negative in this sum rule, indicating that $g_{\pi N}$ is negative.
The sign of this result however depends on the Borel window chosen.
Restricting the Borel window to smaller Borel masses, the extracted
$c$ becomes positive though small in magnitude. 
The slope of the Borel curve is also large, indicating that
there is a large contribution from the undetermined single pole terms.  
The thin dashed curve ( for $S_\pi = 2.57$ GeV$^2$) 
in Fig.~\ref{fig2}~(b) is steeper than the
thick dashed curve (for $S_\pi = 2.07$ GeV$^2$). 
In a sum rule, the larger continuum threshold usually suppresses the
continuum contribution further.
Since a more steeper curve is expected as we further suppress the continuum,
this sum rule contains very large unknown single pole terms.
This provides a very important issue which should be properly addressed 
in the construction of a sum rule.
{\it The unknown single pole terms could be small or large 
depending on a specific sum rule one considers.}    

From the three results, we showed that the extracted parameter, here $c$,
could be totally different depending on how we construct a sum rule.
Even the sign of the parameter is not well fixed. 
Certainly the 
$\gamma_5 \sigma_{\mu\nu} q^\mu p^\nu$ sum rule has nice features,
such as small contributions from the continuum and the unknown single pole.
And when it is combined with the nucleon odd sum rule, it provides
$g_{\pi N}$ reasonably close to its empirical value~\cite{hung}.
But the other sum rules do not provide a reasonable or stable
result.  
It is not clear if this is due to the lack of convergence in the OPE or 
due to the limitations in the sum rule method itself.  To answer such 
questions, it would be useful to analyze the OPE side further.  
However our analysis raises an issue whether or not a sum rule
based on one specific Dirac structure is reliable.

Still, regarding the sensitivity of $S_\pi $ and the unknown single
pole contribution, we can provide a reasonable explanation based on
effective model formalism developed in Section~\ref{sec:effective}.  
Results from the two sum rules,  $i\gamma_5$ and
 $\gamma_5 \sigma_{\mu\nu} q^\mu p^\nu$ structures,
share similar properties. 
As can be seen from Table~\ref{tab},
for the $i\gamma_5$ sum rule, the extracted $c$ is $-0.00033$ GeV$^6$
when $S_\pi =2.07$ 
GeV$^2$ is used.  For $S_\pi =2.57$ GeV$^2$, $c =-0.00016$ GeV$^6$.
So the difference is 0.00017 GeV$^6$.  This difference is close to 
the difference from the $\gamma_5 \sigma_{\mu\nu} q^\mu p^\nu$
case.  Furthermore, the magnitude of $b$ is relatively close in the
two sum rules. 
These common behaviors of the two sum rules 
are expected because, as we briefly mentioned in section~\ref{sec:effective},
their phenomenological structures for the
continuum and the unknown single poles are almost the same except for
the possible small term containing the derivative of the pion form factor.
[See  Eq.~(\ref{shphen1}).]
The similar slope and the similar contribution from $S_\pi$ are
actually related as can be seen from Eq.~(\ref{spec2}).  
In Eq.~(\ref{spec2}), the terms corresponding to the unknown single
poles have the same relative sign between the positive- and
negative-parity resonances  as the terms corresponding to
the continuum.  If we assume that the sign of $\lambda_+ g_{\pi N N_+}$
is opposite to that of $\lambda_- g_{\pi N N_-}$, then there
is a cancellation between the two resonances. 
Thus, with this sign assignment, we expect both terms,
unknown single pole and the ``step-like'' continuum,
contribute less to the sum rules.  This is what Fig.~\ref{fig2}
indicates.
As Eq.~(\ref{spec2}) is independent of the coupling schemes,
this explanation is valid even for the PV case.
The sign assignment, within the PS coupling scheme, also explains
the large slope and strong sensitivity of $S_\pi$ in 
the $i\gamma_5\not\!p$ sum rule.
From Eq.~(\ref{spec1}) with the
sign assignment, 
negative- and positive-parity
resonances add up for the undetermined single pole and the continuum,
yielding large contribution to the two.

This explanation for the $i\gamma_5\not\!p$ sum rule can be changed
for the PV coupling scheme. 
For the case with the undetermined single pole,
as can be seen from Eq.~(\ref{vspec1}), resonances with different
parities cancel each other also for the $i\gamma_5\not\!p$ case under 
the sign assignment introduced above.
However, there is an additional single pole coming from $N \rightarrow N$
which could explain the large slope. Its contribution to
$A$ in Eq.~(\ref{bksum}) can be calculated to be $-1/2m$.  
In terms of magnitude,
it contributes 50\% of the LHS at $M^2\sim 1$ GeV$^2$ with the opposite sign
from the first term.  Since $c$ is negative as we showed in
Table~\ref{tab}, $g_{\pi N}$ is also negative. Since $b \sim g_{\pi N} A$,
the unknown single pole term is positive which can explain the 
large and positive slope in this sum rule. 
As for the continuum,  
Eq.~(\ref{vspec2}) shows that   
there are other contributions
associated with $N^* \rightarrow N^*$ whose magnitudes
can not be estimated.   
Even though we can not say that
the large continuum only comes from adding up the positive- and
negative-parity resonances, this is not contradictory to the
sign assignment for $g_{\pi NN_+} \lambda_+$ and
$g_{\pi NN_-} \lambda_-$.
Note however that the negative sign of $c$ is not firmly established in this
sum rule for the $i\gamma_5\not\!p$ structure as there is a possibility 
that $c$ can be positive for different Borel window chosen.
In this case, the positive and large slope of the Borel curve can
not be well explained within the effective model.

Nevertheless, our study in this work, though it was specific to
the two-point nucleon correlation function with pion,
raises important issues in 
applying QCD sum rules in calculating various physical quantities. 
Most QCD sum rule calculations are performed based
on a specific Dirac structure without justifying the use of
the structure.  As we presented in this work, a sum rule result 
could have a strong dependence on the specific Dirac structure
one considers.  This dependence is  driven by the
way how the sum rule is constructed or by the difference in
the continuum contributions or the unknown single pole terms.   
The continuum and the unknown single pole terms are large in
some case while they are small in other cases.

\section{Summary}
\label{sec:summary}

In this work, we have presented three different sum rules for
the two-point correlation function with pion, 
$i\int d^4x e^{iq\cdot x} \langle 0| T J_N(x) {\bar J}_N(0)|\pi(p)\rangle$,
beyond the soft-pion limit. The PS and PV coupling scheme independence
has been imposed in the construction of the sum rules. 
We have corrected an error in the previous sum rule in Ref.~\cite{krippa}
and found that the sum rule contains
large contribution from the unknown single pole, $b$,
and the continuum.  
On the other hand, the sum rules for
$i\gamma_5$ and 
$\gamma_5 \sigma_{\mu\nu} q^\mu p^\nu$ structures share similar
properties, relatively similar contributions from the continuum and the
unknown single pole.  
By making specific models for higher resonances, we have explained 
how the latter two sum rules are different from
the $i\gamma_5\not\!p$ sum rule. 
Within the PS coupling scheme, the difference can be well explained by
the cancellation or addition of 
the positive- and negative-parity resonances in higher mass states.  
Within the PV coupling scheme,  the large slope of the Borel curve
in the $i\gamma_5\not\!p$ sum rule can be attributed to the single
pole coming from $N\rightarrow N$ transition even though
this explanation is limited to the case with negative value of $g_{\pi N}$.
The value of $c$ extracted from the $i\gamma_5$ and 
$\gamma_5 \sigma_{\mu\nu} q^\mu p^\nu$ sum rules are different.
For the $i\gamma_5$ sum rule, in order to eliminate the coupling
scheme dependence, we need to impose the on-mass-shell condition before
the matching the OPE and phenomenological correlators. Then a significant
cancellation occurs and it makes the $i \gamma_5 $ sum rule less
reliable.
We have stressed that in the construction of a sum rule, a care must be
taken.

\acknowledgments
This work is supported in part by the 
Grant-in-Aid for JSPS fellow, and
the Grant-in-Aid for scientific
research (C) (2) 08640356 
of  the Ministry of Education, Science, Sports and Culture of Japan.
The work of  H. Kim is also supported by Research Fellowships of
the Japan Society for the Promotion of Science.
The work of S. H. Lee is supported by  KOSEF through grant no. 971-0204-017-2
and 976-0200-002-2 and  by the
Korean Ministry of Education through grant no. 98-015-D00061.

\begin{table}
\caption{ 
The best-fit values for the parameters $c$ and $b$ obtained within the
Borel window $0.8 \le M^2 \le 1.2$ GeV$^2$.
The numbers in parenthesis are obtained when $S_{\pi}=2.57 $ GeV$^2$ is used.}

\begin{center}
\begin{tabular}{ccc}
& $c$ (GeV$^6$) & $b$ (GeV$^4$)  \\
\hline\hline
$i\gamma_5\not\!p$ & -0.00022 (-0.0023) & 0.011 (0.0145) \\
$i\gamma_5$ & -0.00033 (-0.00016) & -0.00183 (-0.0021)  \\
$\gamma_5 \sigma_{\mu\nu} q^\mu p^\nu$ & 0.00308 (0.002906) & 0.00257 
(0.0029) \\
\end{tabular}
\end{center}
\label{tab}

\end{table}

\begin{figure}
\caption{ The result of Birse and Krippa's sum rule
for $g_{\pi N}$ is shown in Figure (a).  The solid line is obtained
by eliminating the unknown single pole term using the
differential operator $1 - M^2 \partial/\partial M^2$ and
the dashed line is obtained simply by neglecting the unknown single pole
term.  Figure (b) is similarly obtained  but after correcting
the factors in the treatment of the continuum.}
\label{fig1}
\vspace{10pt}
\caption{(a) Borel curves obtained from  the three different  sum rules.
The thick solid line (thick dashed line)
is for the $\gamma_5 \sigma^{\mu\nu} q_\mu p_\nu$ ($i\gamma_5\not\!p$)    
structure.  The thick dot-dashed line is for the
$i\gamma_5$ structure.   Corresponding thin lines are
obtained when $S_\pi =2.57 $ GeV$^2$ is used. (b) The curves obtained
by linearly fitting the Borel curves within the range, 
$0.8 \le M^2 \le 1.2 $ GeV$^2$. The thin solid lines and thin
dot-dashed lines are almost indistinguishable from the
corresponding thick lines.}

\label{fig2}
\end{figure}
\eject

\setlength{\textwidth}{6.1in}   
\setlength{\textheight}{9.in}  
\begin{figure}
\centerline{%
\vbox to 2.4in{\vss
   \hbox to 3.3in{\includegraphics{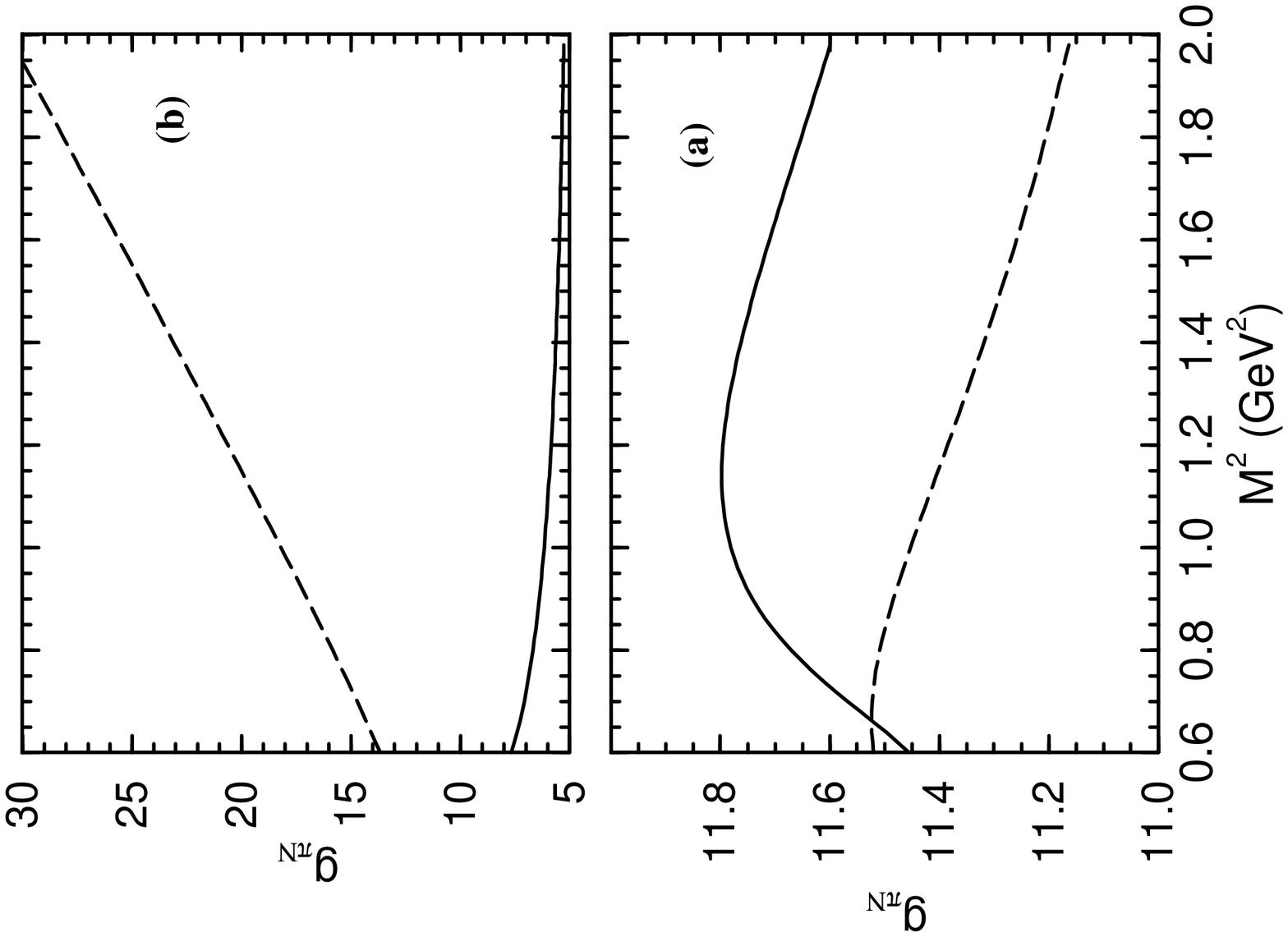}\hss}}
}
\bigskip
\vspace{400pt}
Figure 1
\end{figure}
\eject
\begin{figure}
\centerline{%
\vbox to 2.4in{\vss
   \hbox to 3.3in{\includegraphics{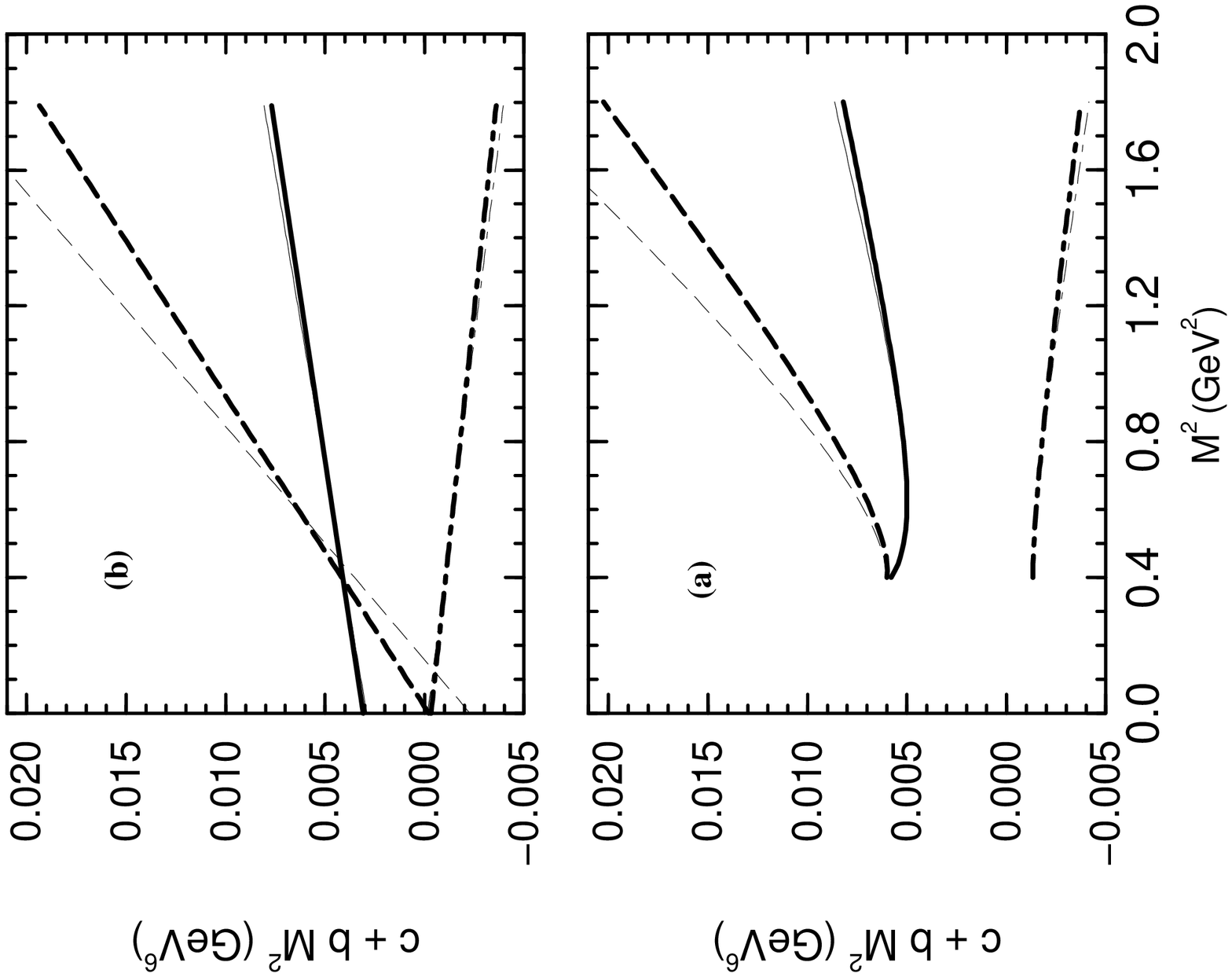}\hss}}
}
\bigskip
\vspace{400pt}
Figure 2
\end{figure}

\begin{references}
\bibitem{SVZ}     {M.A. Shifman, A.I. Vainshtein, and V.I. Zakharov,
                            Nucl. Phys. {\bf B 147} (1979) 385,448.}
\bibitem{qsr}     {For general review see L.J. Reinders, H. Rubinstein and
                    S. Yazaki, Phys. Rep. {\bf 127} (1985) 1.}

\bibitem{hat}     {H. Shiomi and T. Hatsuda,
                            Nucl. Phys. {\bf A 594} (1995) 294.}
\bibitem{krippa}     {M. C. Birse and B. Krippa,
                            Phys. Lett. B {\bf 373} (1996) 9;
                            Phys. Rev. C {\bf 54} (1996) 3240.}
\bibitem{bely1}   {V. M. Belyaev and Ya. I. Kogan,
                   JETP Lett. {\bf 37}, (1983) 730;
                   B. L. Ioffe and A. G. Oganesian,
                   Phys. Rev. D {\bf 57} (1998) R6590.}
\bibitem{hung}     {Hungchong Kim, Su Houng Lee and Makoto Oka,
                    {\it Los Alamos Preprint nucl-th/9809004},
                            {\it submitted to Physics Letters B}.}
\bibitem{jin2}     {X. Jin and J. Tang,
                            Phys. Rev. D {\bf 56} (1997) 515.}
\bibitem{ioffe1}     {B. L. Ioffe,
                            Nucl. Phys.  {\bf B188} (1981) 317.}
 
\bibitem{bely}     {V. M. Belyaev, V. M. Braun, A. Khodjamirian and R. R\"uckl,
                            Phys. Rev. D {\bf 51} (1995) 6177.}
\bibitem{ioffe2}     {B. L. Ioffe and A. V. Smilga,
                            Nucl. Phys. {\bf B 232} (1984) 109.}
\bibitem{nov}     {V. A. Novikov, M. A. Shifman, A. I. Vainshtein, 
                   M. B. Voloshin and V. I. Zakharov, 
                            Nucl. Phys. {\bf B 237} (1984) 525.}
\bibitem{Ovc}       {A. A. Ovchinnikov and A. A. Pivovarov,
                            Yad. Fiz. {\bf 48} (1988) 1135.}

\end{references}
\end{document}